\documentclass[twocolumn,aps,showpacs,floatfix,prc]{revtex4}
\usepackage[dvips]{epsfig}

\begin{document}
 
\title{Hydrodynamical model for $J/\psi$ suppression and elliptic flow}
\author{A. K. Chaudhuri}
\email[E-mail:]{akc@veccal.ernet.in}
\affiliation{Variable Energy Cyclotron Centre,\\ 1/AF, Bidhan Nagar,
Kolkata 700~064, India}

\begin{abstract}
In a hydrodynamic model, we have studied $J/\psi$ suppression and elliptic flow in Au+Au collisions at RHIC energy $\sqrt{s}$=200 GeV. 
At the initial time, $J/\psi$'s are randomly distributed in the fluid.  As the fluid evolve in time, the free streaming $J/\psi$'s are dissolved if the local fluid  temperature exceeds a melting temperature $T_{J/\psi}$. 
Sequential melting of charmonium states
($\chi_c$, $\psi\prime$ and $J/\psi$), with melting temperatures $T_{\chi_c}=T_{\psi\prime} \approx 1.2T_c$, $T_{J/\psi} \approx2T_c$ and  feed-down fraction $F\approx 0.3$, is consistent with the PHENIX data  on
$J/\psi$ suppression and near zero elliptic flow for $J/\psi$'s. It is also shown that the model will require substantial regeneration of charmoniums, if the charmonium states dissolve at temperature close to the critical temperature, $T_{\chi_c}=T_{\psi\prime} \leq T_c$, $T_{J/\psi}\approx1.2T_c$. The regenerated charmoniums will have positive elliptic flow.
 \end{abstract}  
\pacs{PACS numbers: 25.75.-q, 25.75.Dw}

\maketitle
 
Matsui and Satz \cite{Matsui:1986dk},  
predicted that in presence of quark-gluon plasma  (QGP), due to color screening, binding
of a $c\bar{c}$  pair  into  a  $J/\psi$  meson will be hindered,
leading to the  so  called  $J/\psi$  suppression  in  heavy  ion
collisions. 
Over  the  years,  several  groups have
measured the $J/\psi$ yield in heavy ion collisions (for a review
of the data prior to RHIC energy collisions, and the interpretations see Refs.  \cite{Vogt:1999cu,ge99}).
In  brief,  experimental  data do show suppression. However, this
could be attributed to the conventional nuclear absorption,  also
present in $pA$ collisions. At RHIC energy ($\sqrt{s}$=200 GeV), PHENIX  collaboration  has made systematic measurements of $J/\psi$ production in nuclear collisions. 
They have measured $J/\psi$ yield in p+p collisions at RHIC and obtained the  reference for the basic invariant yield \cite{Adler:2003qs,Adler:2005ph,Adare:2006kf}. Measurements of $J/\psi$ production  in d+Au collisions
\cite{Adler:2005ph,Adare:2007gn} give reference for cold nuclear matter effects. $J/\psi$ production in d+Au collisions are consistent with cold nuclear matter effect quantified in a Glauber model of nuclear absorption with $\sigma_{abs}=2\pm 1$ mb \cite{Vogt:2005ia}.
Cold and hot nuclear matter effects are studied  in
Au+Au and Cu+Cu collisions, where yields are measured as a function of collision centrality \cite{Adler:2003rc,Adare:2006ns,Adare:2008sh}. 
At RHIC energy, it has been
argued that rather than suppression, charmonium's will be enhanced
\cite{ Thews:2000rj,Braun-Munzinger:2000px}. 
Due to large initial energy, large number of $c\bar{c}$ pairs will be
produced in initial hard scatterings. Recombination of $c\bar{c}$
can occur enhancing the charmonium production. Apparently, both the PHENIX data on
$J/\psi$ production in Au+Au and in Cu+Cu collisions, are not consistent
with models which predict $J/\psi$ enhancement
\cite{ Thews:2000rj,Braun-Munzinger:2000px}.   $J/\psi$'s are suppressed both in Au+Au and Cu+Cu collisions, suppression is more in central collisions than in peripheral collisions.

Recently PHENIX collaboration    \cite{Silvestre:2008tw} measured the elliptic flow for $J/\psi$ in 20-60\% Au+Au collisions. 
Elliptic flow for $J/\psi$ is an important observable. It can
test whether or not the $J/\psi$ production is dominated by recombination. In p+p and Au+Au collisions, PHENIX has measured semi-leptonic decay electrons from heavy quarks \cite{Hornback:2008ur}.  In Au+Au collisions decay electrons has positive elliptic flow. If recombination of $c\bar{c}$ is a major source of $J/\psi$ in Au+Au collisions, as suggested in \cite{ Thews:2000rj,Braun-Munzinger:2000px},   $J/\psi$'s will inherit some of their flow.  The PHENIX measurements of $J/\psi$ elliptic flow has large error bars. Integrated $v_2$ is consistent with zero, $v_2=-0.10 \pm 0.10\pm 0.02$.  

Recently, in a hydrodynamic based model \cite{Gunji:2007uy, Chaudhuri:2008qq,Chaudhuri:2008if}, centrality dependence of $J/\psi$ suppression in Au+Au and Cu+Cu collisions are studied. In the model, $J/\psi$'s are randomly produced in initial NN collisions. As the fluid evolve, free streaming $J/\psi$'s are melted if the local fluid temperature exceed a threshold temperature $T_{J/\psi}$. Sequential melting of charmonium states
($\chi_c$, $\psi\prime$ and $J/\psi$), with melting temperatures $T_{\chi_c}=T_{\psi\prime} \approx 1.2T_c$, $T_{J/\psi} \approx2T_c$ and  feed-down fraction $F\approx 0.3$,  explains the 
PHENIX data  on the centrality dependence of  $J/\psi$ suppression in   Au+Au collisions. $J/\psi$ $p_T$ spectra and the nuclear modification factor in  Au+Au collisions are also well explained in the model. The model leaves  little or no room for   regeneration of $J/\psi$ due to recombination of $c\bar{c}$ pairs, as suggested in \cite{ Thews:2000rj,Braun-Munzinger:2000px}. The model is also consistent with zero elliptic flow for $J/\psi$'s. Initially $J/\psi$'s are produced randomly, they do not have any flow. In later times also, the free streaming $J/\psi$'s can not acquire any flow. 

The dissociation or melting temperatures of different charmonium states obtained in \cite{Gunji:2007uy, Chaudhuri:2008qq,Chaudhuri:2008if} are in agreement with potential model calculations for charmonium states at finite temperature \cite{Satz:2006kb}.
 Due to heavy mass of charm quarks, charmonium states can be studied in non-relativistic potential models. Ground state properties of charmonium stated are well explained using the Cornell potential, $v(r)=\sigma r -\frac{\alpha}{r}$, with string tension $\sigma \approx $0.2 $GeV^2$ and gauge coupling $\alpha\approx \pi/12$ \cite{Satz:2006kb}. Lattice QCD can  provide for the heavy quark potential at finite temperature. Finite temperature potential models  
 indicate that charmonium states, $J/\psi$(1S), $\chi_c$(1P) and $\psi\prime(2S)$ dissociate respectively at temperatures,
$T_{J/\psi}\approx 2 T_c$, $T_{\chi_c}=T_{\psi\prime} \approx 1-1.2 T_c$ \cite{Satz:2006kb}.   However, very 
  recently, in  \cite{Mocsy:2007jz,Mocsy:2008zz}  
 quarkonia spectral function in QGP is determined using a potential motivated by lattice QCD results on free energy of static quark-antiquark pair. Surprisingly, charmoniums are found to melt at much lower temperature,   $T_{J/\psi}\approx 1.2T_c$, $T_{\chi_c}=T_{\psi\prime} \leq 1.2 T_c$.

In the present paper, we show that if the charmonium states dissolve at temperature close to the critical temperature, the PHENIX data on $J/\psi$ suppression in Au+Au collisions are not explained in the hydrodynamic model \cite{Gunji:2007uy, Chaudhuri:2008qq,Chaudhuri:2008if}. $J/\psi$'s will be more suppressed than in the experiment.  
Data on $J/\psi$ suppression can only be explained with substantial regeneration of $J/\psi$ during the evolution. It is also shown that regenerated charmoniums will have positive elliptic flow. Conversely, small positive elliptic flow for $J/\psi$ will indicate regeneration of charmoniums in RHIC energy collisions.

\begin{figure}[t]
 \center
 \resizebox{0.35\textwidth}{!}
 {\includegraphics{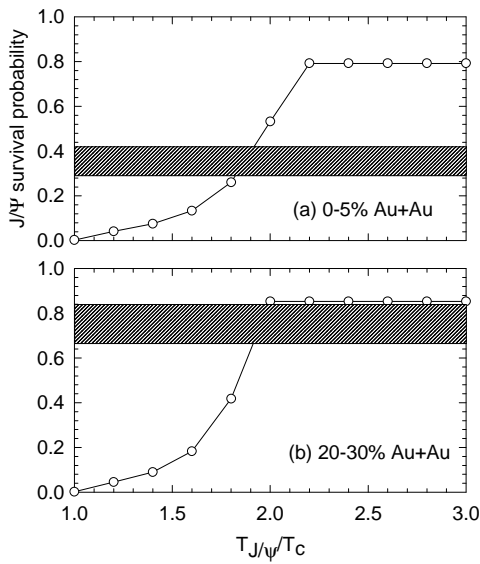}}
\caption{The open circles are hydrodynamic model predictions for $J/\psi$ survival probability in (a) 0-5\%   and (b) 20-30\%  centrality Au+Au collisions,  as a function of $J/\psi$ melting temperature 
($T_{J/\psi}$). The shaded regions in panel (a) and (b) indicates PHENIX measurements for $J/\psi$ survival probability  \cite{Adare:2006ns,Leitch:2007wa,Gunji:2007fi}.
Note that if $J/\psi$ dissociation temperature is $T_{J/\psi} \approx 1.2T_c$, experimental data will be under explained.
}\label{F1}
\end{figure}
 
Details of the hydrodynamic model used here can be 
found in \cite{Chaudhuri:2008qq,Chaudhuri:2008if}. Briefly,
it is assumed that 
in high energy nuclear collisions, a deconfined phase (QGP) is produced, which expands, cools, undergoes 1st order phase transition to hadronic fluid at the critical temperature ($T_c$=164 MeV) and then further cools to freeze-out at temperature $T_F$=130 MeV.  
Assuming longitudinal boost-invariance, the space-time evolution of the fluid is obtained by solving the energy-momentum conservation equation $\partial_\mu T^{\mu\nu}=0$ , with initial conditions as determined in \cite{QGP3}, e.g. initial time $\tau_i$=0.6 fm, initial central entropy density $S_{ini}$=110 $fm^{-3}$, corresponding to energy density $\varepsilon_i\approx$30 $GeV/fm^3$. To obtain the survival probability of $J/\psi$'s in
an expanding medium, we proceed as follows:
at the initial time $\tau_i=0.6 fm/c$,    we  randomly distribute a fixed number of $J/\psi$'s in the transverse plane. They are assumed to be free streaming   unless dissolved in the medium. 
Each $J/\psi$ is characterised by 4 random numbers. Two random numbers ($R_1,R_2$) indicate its transverse position (${\bf r}_\perp$),
and two random numbers ($R_3,R_4$) its transverse momentum $\vec{p}_T$. 
Random numbers $R_1$ and $R_2$ are distributed according to the transverse profile of the number of binary collisions ($N_{coll}$), calculated in a Glauber model. Random numbers $R_3$ is distributed according the
 power law \cite{Yoh:1978id},

\begin{equation}\label{eq2}
B\frac{d\sigma}{dyd^2p_T}=\frac{A}{[1+(p_T/B)^2]^6} (nb/GeV^2),
\end{equation}

\noindent with $A=4.23$ and $B=4.1$, which well describe the invariant distribution of measured $J/\psi$'s in p+p collisions .   The random number $R_4$ is distributed uniformly within [0-2$\pi$].  
 
The survival probability of a $J/\psi$ inside the expanding QGP is calculated as \cite{Gunji:2007uy},

\begin{equation} \label{eq3}
S_{J/\psi}(\tau)=exp\left[-\int_{\tau_i}^\tau \Gamma_{dis}(T({\bf r}_{\perp}(\tau^\prime))) d\tau^\prime \right]
\end{equation}

\noindent where $T({\bf r}_\perp)$ is the temperature of the fluid at the transverse position $r_\perp$, $\Gamma_{dis}(T)$ is the decay width of $J/\psi$ at temperature $T$. $\tau_i$ is the initial time for hydrodynamic evolution. We continue the evolution till the freeze-out time $\tau_F$, corresponding freeze-out temperature $T_F$=130 MeV. For the decay width $\Gamma_{dis}$ we use,

\begin{eqnarray} \nonumber
\Gamma_{dis}(T) &= & \infty; \hspace{1cm} T>T_{J/\psi} \\
\Gamma_{dis}(T) &= &\alpha (T/T_c -1)^2; \hspace{1cm}T<T_{J/\psi} \label{eq5}
\end{eqnarray}

In Eq.\ref{eq5}, $\alpha$  is the thermal width of the state at $T/T_c$=2. NLO perturbative calculations suggest that
$\alpha > 0.4 GeV$ \cite{Park:2007zza}. Presently we use $\alpha$=0.4 \cite{Chaudhuri:2008if}.

In Fig.\ref{F1}(a) and (b), hydrodynamic model predictions for $J/\psi$ survival probability ($S_{J/\psi}$), in 0-5\% and 20-30\% centrality Au+Au collisions are shown as a function of the melting temperature. As expected,   survival probability increases as the melting temperature increase. Survival probability continue to increase till $T_{J/\psi}$ exceed the peak temperature of the fluid, beyond which, even if $T_{J/\psi}$ is increases, survival probability remain unchanged. 
The shaded region
in Fig.\ref{F1}(a) and (b) represent the experimental survival probability as determined in the PHENIX experiment \cite{Adare:2006ns}. The cold nuclear matter effect is subtracted out \cite{Adare:2006ns,Leitch:2007wa,Gunji:2007fi}. It is
evident from Fig.\ref{F1} 
that if $J/\psi$'s dissolve at temperatures close to the critical temperature, $T_{J/\psi}\approx 1.2 T_c$, as 
predicted in recent calculations \cite{Mocsy:2007jz,Mocsy:2008zz},
PHENIX data on $J/\psi$ suppression is not explained in the model.
 Most of the initially produced $J/\psi$'s ($\sim$ 98\%) are dissolved in the medium. The data could only be explained, if during the evolution, $c\bar{c}$ pairs recombine to regenerate $J/\psi$. In the discussion, we have neglected the effect of higher states $\chi_c$ and $\psi\prime$. Nearly 30-40\% of observed $J/\psi$'s are from decay of the higher states. However, since $\chi_c$ and $\psi\prime$ melt at temperature lower than that for $J/\psi$'s, their inclusion will require further regeneration of charmoniums. If on the other hand, $J/\psi$'s do survive high temperature (e.g. $T_{J/\psi}\approx 2T_c$), experimental data is explained in the model, without any need of recombination of $c\bar{c}$ pairs. Indeed, as shown in \cite{Chaudhuri:2008if}, PHENIX data on centrality dependence of $J/\psi$ suppression are well explained in the model with sequential melting of charmoniums states with melting temperatures, $T_{\chi_c}=T_{\psi\prime} \approx 1.2T_c$, $T_{J/\psi} \approx2T_c$ and  feed-down fraction $F\approx 0.3$, with little or no scope for recombination of $c\bar{c}$ pairs and regeneration of charmoniums in the QGP phase. 

\begin{figure}[t]
 \center
 \resizebox{0.35\textwidth}{!}
 {\includegraphics{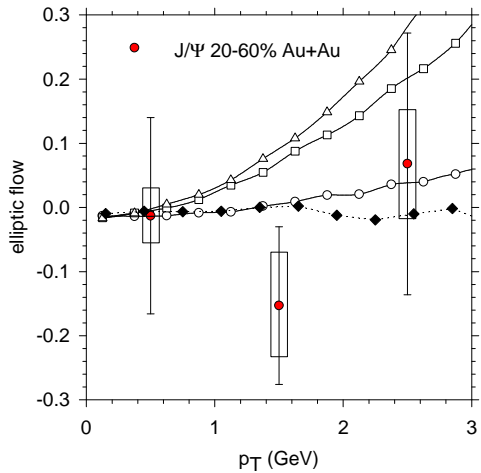}}
\caption{  Black diamonds are hydrodynamic model \cite{Chaudhuri:2008if} predictions
for $J/\psi$ elliptic flow in 20-60\% Au+Au collisions. Elliptic flow for $J/\psi$'s,   randomly generated according to the thermal distribution in Eq.\ref{eq4}, in 0-5\% and 20-30\% and 20-60\% Au+Au collisions  are shown as
blank circles, squares and  triangles. The black circles are PHENIX measurements \cite{Silvestre:2008tw} 
for $J/\psi$  elliptic flow  in 20-60\% Au+Au collisions .
  }\label{F2}
\end{figure}

Hydrodynamic model of $J/\psi$ suppression thus indicate that if the charmonium states dissolve at temperature close to critical temperature, more than $\sim$ 98\% of observed $J/\psi$ are regenerated from $c{\bar c}$ recombination during the QGP phase. If on the other hand, charmonium ground state survive high temperature,
regeneration of charmoniums will be minimum.
Regenerated charmoniums can have positive elliptic flow. As mentioned earlier,
PHENIX collaboration measured single electrons from the semi-leptonic decay of heavy flavors in p+p and Au+Au collisions \cite{Hornback:2008ur}. Semi-leptonic decay electrons in minimum bias Au+Au collisions have positive elliptic flow, indicating positive elliptic flow for the parent heavy quarks. 
If $J/\psi$'s are produced from $c\bar{c}$ recombination in the QGP phase, they could inherit their flow. To obtain an idea about the elliptic flow for $J/\psi$, due to recombination, we randomly generate $J/\psi$'s at the critical 
temperature $T_c$ according to the equilibrium distribution,

\begin{equation} \label{eq4}
E\frac{dN}{d^3p} \propto exp(-p.u(x)/T_c)  
\end{equation}

\noindent where $u(x)$ is fluid velocity at the freeze-out surface at $T=T_c$.
Eq.\ref{eq4} is certainly an assumption.  $J/\psi$'s can be regenerated throughout the deconfined phase. However, Eq.\ref{eq4} suffice to give the general idea of  flow for the regenerated $J/\psi$'s.  
 Fluid velocity distribution at the critical temperature $T_c$ can be obtained from the hydrodynamical model of evolution.
In Fig.\ref{F2}, differential elliptic flow for $J/\psi$'s, randomly generated according to  the distribution in Eq.\ref{eq4}, in 0-5\% and 20-30\% and 20-60\% Au+Au collisions are shown. While in central (0-5\%) collisions, elliptic flow is small, in peripheral   
(20-30\% or in 20-60\%) Au+Au collisions, regenerated charmoniums have substantial elliptic flow. For example $J/\psi$'s produced randomly according to the distribution Eq.\ref{eq4} have $\sim$ 15-20\% $v_2$ at $p_T$=2 GeV. 
In Fig.\ref{F2},  black circles are PHENIX measurements for $J/\psi$ elliptic flow in 20-60\% Au+Au collisions. The error bars are large and definitive conclusion about $J/\psi$ elliptic flow can not be drawn. Apparently, regenerated charmions have more elliptic flow than in experiment.
In Fig.\ref{F2}, we have also shown the elliptic flow for $J/\psi$ in the hydrodynamic model \cite{Chaudhuri:2008if}, 
with melting temperatures,
$T_{J/\psi}=2 T_c$, $T_{\chi_c}=T_{\psi\prime}=1.2T_c$ and feed down fraction F=0.3 (the black circles). As stated earlier, the model is consistent with PHENIX data on $J/\psi$ suppression without any recombination. The elliptic flow in the hydrodynamic model is consistent with zero flow.
It is expected also. In the model, $J/\psi$'s are randomly generated   within the angular range ($0-2\pi$). Initially they have zero elliptic flow. During the evolution also, free streaming $J/\psi$'s do not acquire  flow. 

Before we summarise, we note that it is possible that difference between elliptic flow of regenerated $J/\psi$'s and $J/\psi$'s surviving the hydrodynamic evolution, may not be as large as depicted in Fig.\ref{F2}. We have assumed $J/\psi$'s are free streaming.  $J/\psi$'s are formed early in the collision.
The formation process may induce elliptic flow. Then elliptic flow of $J/\psi$'s surviving hydrodynamical evolution will be larger than obtained presently with the assumption of free steaming. It is difficult to ascertain the induced elliptic flow in initial $J/\psi$'s and uncertainty at the present evaluation
can not be quantified. Present model, with the assumption of free streaming give the lower limit of elliptic flow of  $J/\psi$'s surviving the hydrodynamic evolution.  Also it is  unlikely that regenerated $J/\psi$ will be 
equilibrated at $T_c$. 
$J/\psi$'s are massive. Their mass is much larger than $T_c$. Elliptic flow of non-equilibrated or partially equilibrated $J/\psi$ will be less than that for completely equilibrated $J/\psi$'s.   Present evaluation, with the assumption of equilibration then give the upper limit of elliptic flow of regenerated $J/\psi$'s.

To summarise, in a hydrodynamical model,   we have 
studied $J/\psi$ suppression in Au+Au collisions at RHIC. In the model, 
at the initial time, $J/\psi$'s are randomly distributed in the fluid.  As the fluid evolve in time, the free streaming $J/\psi$'s are dissolved if the local fluid  temperature exceeds a threshold temperature $T_{J/\psi}$.   It is
shown that if, as suggested in some recent works \cite{Mocsy:2007jz,Mocsy:2008zz},
$J/\psi$ do not survive much above the critical temperature, PHENIX data on $J/\psi$ suppression are not explained in the model. Most of the initially produced $J/\psi$'s are dissolved in the medium and the model predictions severely underestimate the PHENIX data on $J/\psi$ suppression in Au+Au collisions. The data will demand recombination of $c\bar{c}$ pairs in the QGP phase. In a simple model, we have also shown that the regenerated $J/\psi$ will have positive elliptic flow. On the other hand, if $J/\psi$'s can survive at high temperature, (e.g. $T_ {J/\psi}\approx 2T_c$), sequential melting of charmonium states
($\chi_c$, $\psi\prime$ and $J/\psi$), with melting temperatures $T_{\chi_c}=T_{\psi\prime} \approx 1.2T_c$, $T_{J/\psi} \approx2T_c$ and  feed-down fraction $F\approx 0.3$, is consistent with PHENIX data on $J/\psi$ suppression. There is no scope for recombination of $c\bar{c}$ pairs. The model also predict zero elliptic flow for $J/\psi$'s.

\end{document}